\documentclass[10 pt]{iopart}

\usepackage{caption2}
\usepackage{graphicx}

\newcommand{\be}{\begin{equation}}
\newcommand{\ee}{\end{equation}}
\newcommand{\ber}{\begin{eqnarray}}
\newcommand{\eer}{\end{eqnarray}}
\newcommand{\de}{\end{equation*}}
\newcommand{\cer}{\begin{eqnarray*}}
\newcommand{\der}{\end{eqnarray*}}

\begin{document}
\title{A time frequency analysis of wave packet fractional revivals}

\author{Suranjana Ghosh and J Banerji}

\address {Theoretical Physics Division
\\ Physical Research Laboratory, Navrangpura, Ahmedabad 380 009, India}

\ead{sanjana@prl.res.in, jay@prl.res.in}

\begin{abstract}
We show that the time frequency analysis of the autocorrelation
function is, in many ways, a more appropriate tool to resolve
fractional revivals of a wave packet than the usual time domain
analysis. This advantage is crucial in reconstructing the initial
state of the wave packet when its coherent structure is
short-lived and decays before it is fully revived. Our
calculations are based on the model example of fractional revivals
in a Rydberg wave packet of circular states. We end by providing
an analytical investigation which fully agrees with our numerical
observations on the utility of time-frequency analysis in the
study of wave packet fractional revivals.

(Figures in this article are in colour only in the electronic
version)
\end{abstract}

\maketitle

\section{Introduction}
In certain quantum systems with nonlinear energy spectra, a
suitably prepared wave packet will, in the course of its
evolution, regain its initial form periodically. This is known as
the revival of the wave packet. At some intermediate times, the
evolving wave packet will break up into a set of replicas of its
original form. This is known as the fractional revivals of the
wave packet \cite{aver,robinett}. It has been shown that the
phenomena of revival and fractional revival occur in the wave
packet dyanmics of various atomic, molecular and optical systems
such as Rydberg
atoms~\cite{rydberg1,rydberg2,rydberg3,rydberg4,rydberg5,rydberg6,jyeazell,stroud,meacher},
optical parametric oscillators~\cite{opo1,opo2,opo3}, the
Jaynes-Cummings model~\cite{jcm1,jcm2,jcm3}, transient signals
from multilevel quantum systems~\cite{leichtle}, potential
wells~\cite{box1,box2,posch1,posch2} and  molecular vibrational
states~\cite{morse,eryo,stolow,ghosh}. Extensions to systems for
which the energy spectrum depends on two quantum numbers, have
also been made in recent years ~\cite{b1,b2,b3,b4,bluhm1,bluhm2}.
These phenomena have been experimentally observed in both atomic
\cite{rydberg4,jyeazell,stroud,meacher} and molecular
 systems \cite{stolow}.

A widely used method for probing the revival dynamics of wave
packets is based on a study of the autocorrelation function
\cite{rydberg2, eryo, jpbnew}. This method is directly related to
the observable signal in the pump-probe type experiments for
studying the wave packet dynamics. The autocorrelation function
for an evolving wave packet $\Psi(r,t)$ is given by the overlap
integral $A(t)=\langle\Psi(r,t)\vert\Psi(r,0)\rangle$.

The auto-correlation function is a time series whose Fourier
Transform (FT) will reveal all the frequencies, but will be unable
to provide any information on when a particular frequency appears.
This is important in the present context as the frequencies are
time-varying. In other words, fractional revivals of a particular
order occurs at particular instants of time. What is really
desirable is a time-frequency analysis of the auto-correlation
function such that we not only know all the frequencies, but also
get information on when a particular frequency occurs. This is the
objective of the present work.

For the purpose of time-frequency analysis, short-time Fourier
transform (STFT) has often been used in the literature. This
method divides the whole time series in several windows, each of
certain fixed width. Then the FT is performed in each window for
obtaining the frequency information. Unfortunately, the
time-frequency information obtained by this method has not always
been satisfactory as its fixed window length does compromise on
the frequency resolution.  The method of continuous wavelet
transform (CWT) \cite{dau,chui,robi,fox,ajp1,ajp2,ajp3} overcomes
the preset resolution problem of STFT by using a variable length
window. This transform, by design provides good localisation in
both time and frequency. The subject area of wavelets, developed
mostly over the last fifteen years, is at the forefront of much
current research in pure and applied mathematics, physics,
computer science and engineering. This transform has emerged over
recent years as a powerful time-frequency analysis. A narrow
window is used for the analysis of the high frequencies and gives
a better time resolution. A wider window is used for the analysis
of low frequencies and gives a better frequency resolution. The
continuous wavelet transform (CWT)
 of a signal $f(t)$ is defined as
\begin{equation}
T(s,\tau)=\frac{1}{\sqrt{s}}\int
f(t)\phi^{*}\left(\frac{t-\tau}{s}\right)dt
\end{equation}
This transformed signal is a function of two variables, $s$ and
$\tau$ that are used respectively to scale and translate the
wavelet window whereas $\phi^{*}$ is the complex conjugate of the
transforming function known as the mother wavelet for the CWT. In
our study we have used the Morlet wavelet as the mother wavelet.
The contribution to the signal energy at the specific scale $s$
 and location $\tau$ is given by the two dimensional wavelet energy
density function known as the scalogram:
\begin{equation}
E(s,\tau)=|T(s,\tau)|^{2}.
\end{equation}
The frequencies are inversely proportional to the scale parameter,
and thus $s$ and $\tau$ together help provide information in the
time-frequency plane.
In this paper, we present a case study to show that a wavelet
based time-frequency analysis is superior in many ways to the
standard time-domain analysis of the auto-correlation function.

\section{Fractional Revivals of a Rydberg wave packet}

We consider a Rydberg wave packet which is a superposition of
circular Hydrogenic states having $l=m=n-1$ \cite{rydberg3}. The
time dependent wave function for a localized wave packet formed as
a one dimensional superposition of eigenstates may be written as

\begin{equation}
\psi(\vec{r},t)=\sum_{n} c_{n} \psi_{n}(\vec{r}) e^{-iE_{n}t}
\end{equation}

 As a pre-requisite for obtaining fractional revivals, we assume that the weighting
 probabilities $|c_{n}|^{2}$ are
 strongly peaked around a mean value $\bar{n}$ with a spread $\Delta n=n_{\mathrm{max}}-n_{\mathrm{min}}\ll \bar{n}$. This allows us to expand the energy eigenvalues
$E_{n}=-(2n^2)^{-1}\label{energy1}$ (in atomic unit) in a Taylor
series in n as follows.
\begin{equation} E_{n}=
E_{\bar{n}}+E'_{\bar{n}}(n-\bar{n})+\frac{1}{2}E''_{\bar{n}}
(n-\bar{n})^2+\frac{1}{6}E_{\bar{n}}(n-\bar{n})^3+....
\end{equation}

Neglecting the overall time dependent phase and considering up to
the  second order term, we may write $E_n$ as
\begin{equation}
E_n=2\pi\left\{\frac{(n-\bar{n})}{T_{\mathrm{cl}}}+\frac{(n-\bar{n})^2}
{T_{\mathrm{rev}}}\right\},\label{phase1}
\end{equation}

where each term in the expansion defines important characteristic
time scale that depend on $\bar{n}$ ;

\begin{equation}
T_{\mathrm{cl}}=\frac{2\pi}{|E'_{\bar{n}}|} ,\;\;\;
T_{\mathrm{rev}}=\frac{2\pi}{\frac{1}{2}|E''_{\bar{n}}|}.\label{timescales}
\end{equation}
Since the energy spectrum is known in this case, one obtains
\begin{equation}
T_{\mathrm{cl}}=2\pi {\bar{n}}^3,\;\;\; T_{\mathrm{rev}}=2
T_{\mathrm{cl}} {\bar{n}}/3.\label{timescales1}
\end{equation}
The absolute square of autocorrelation function is
\begin{equation}
f(t)=|A(t)|^{2}=\sum_{n,m}|c_{n}|^{2} |c_{m}|^{2} e^{-i
E_{nm}t/\hbar},\label{auto20}
\end{equation}
where $E _{nm}=E_{n}-E_{m}$.

\begin{figure}[htbp]
  \includegraphics[width=2.6in,height=1.94in]{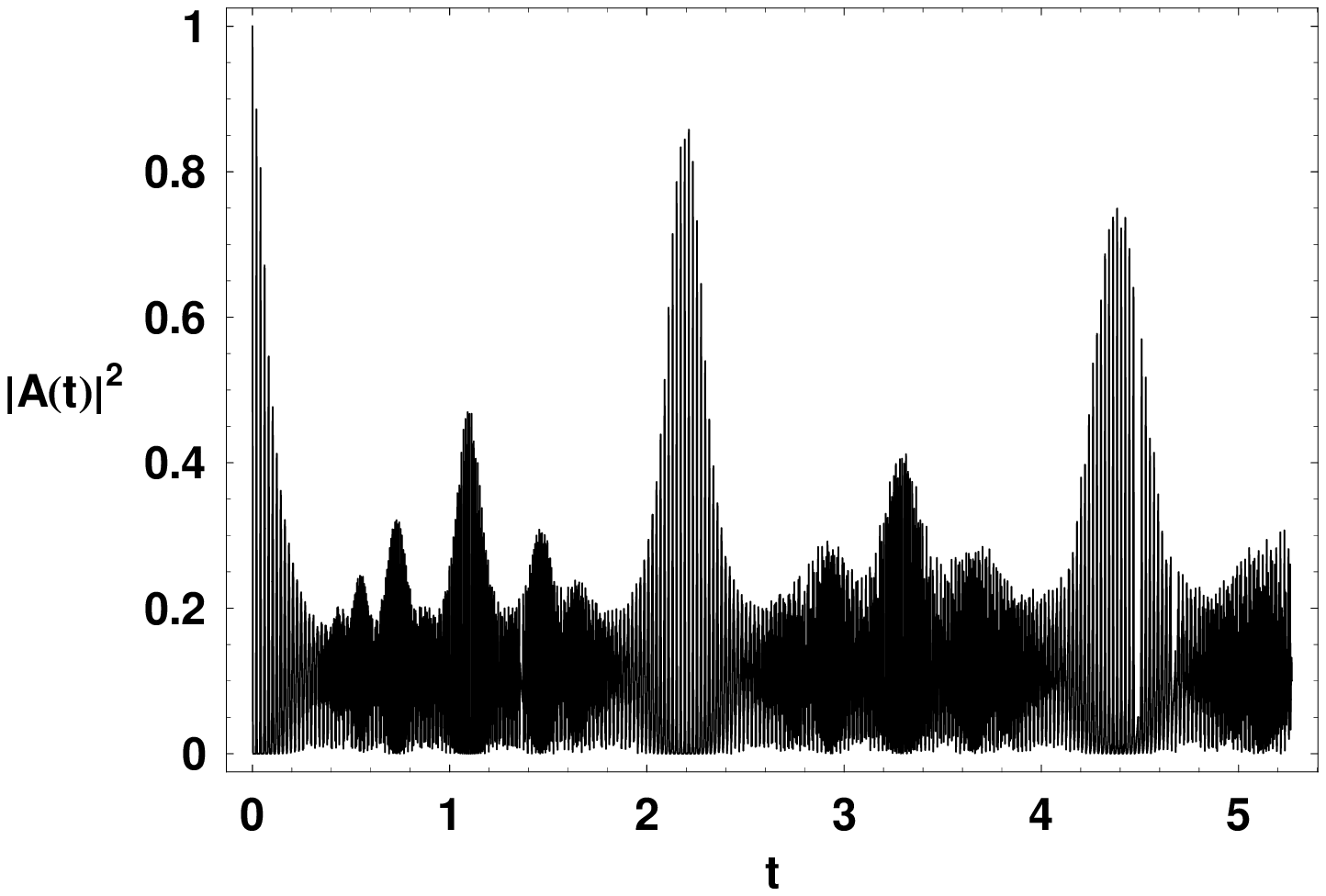}\;\;\;
  \includegraphics[width=2.25in,height=1.86in]{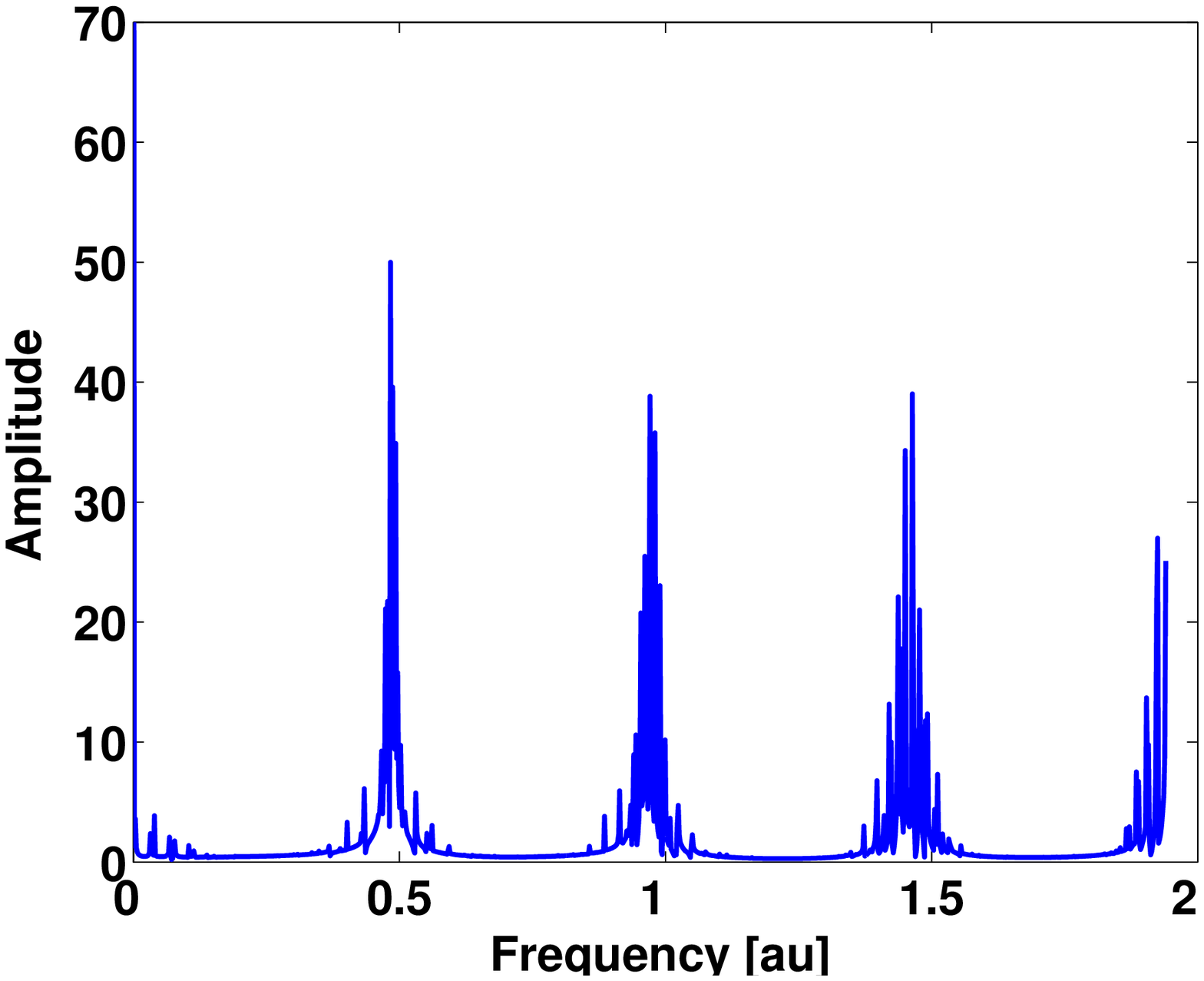}
\caption{(a) Autocorrelation function of a Rydberg wave packet
with $\bar{n}=320$, $\Delta n=40$ and $\sigma=2.5$. Time $t$ is in
a.u. (in the unit of $10^{10})$. (b) Spectral components present
in the autocorrelation function of the Rydberg wave packet.
Frequency is in a.u (in the unit of $10^{-8}$).} \label{auto1}
\end{figure}

Fig.~\ref{auto1}(a) shows a plot of $|A(t)|^{2}$ as a function of
time. This plot was generated by choosing  $|c_{n}|^{2}$ as a
Gaussian distribution with  $\bar{n}=320$, $\Delta n=40$ and a
FWHM given by $\sigma=2.5$.

Peaks appearing in Fig.~\ref{auto1} (a), are the signature of
revivals and fractional revivals. Fast Fourier transform (FFT) of
this time series data gives the individual spectral components in
the frequency plane as shown in Fig.~\ref{auto1} (b). Although we
can get complete frequency information in this way, we do not have
any idea on which frequency appears at what time. On the other
hand, the autocorrelation time series can be recovered by using
the inverse FFT, but the frequency information goes away
completely.

Our goal is to acquire some frequency information in some
particular times of interest. In the next section, we make use of
continuous wavelet transform to investigate how one is able to
resolve time and frequency in a better way.

\section{Time-Frequency Analysis}

\begin{figure}[htbp]
\centering
\includegraphics[width=3.2in,height=2.2in]{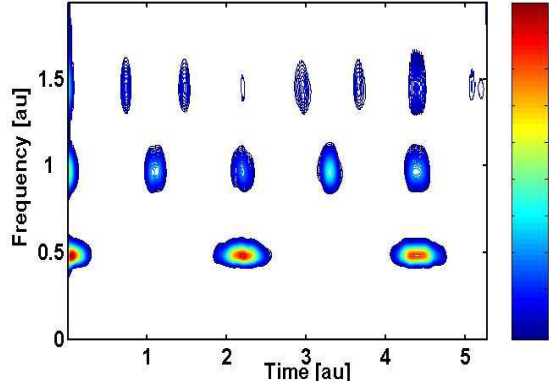}
\caption{Time-frequency representation of the autocorrelation
function of Rydberg atom. Time $t$ (in the unit of $10^{10}$) and
frequency (in the unit of $10^{-8}$) are in a.u.}\label{ryd3}
\end{figure}

Wavelet based time-frequency representation or scalogram of the
time series data $f(t)=|A(t)|^{2}$  is shown in Fig.~\ref{ryd3}.
The scalogram was computed by using the {\it Time-Frequency Tool
Box (TFTB) for MatLab} \cite{toolbox}. Here, we have used the
Morlet wavelet, described as a complex exponential modulated by a
Gaussian envelope. It is a function of time and given by
$\varphi(t)=\pi^{-1/4}e^{i\omega_{0}t}e^{-t^{2}/2}$, with the
central frequency $\omega_{0}$. Let $\Delta t$ and $\Delta\omega$
be the RMS duration and bandwidth respectively of the mother
wavelet $\varphi(t)$, where $\Delta t$ is given by
\begin{equation}\label{delt}
\Delta
t\equiv\sqrt{\frac{\int_{-\infty}^{\infty}(t-t_{0})^{2}|\varphi(t)|^{2}
dt}{\int_{-\infty}^{\infty}|\varphi(t)|^{2} dt}},
\end{equation}
The term inside the square root is the second moment of the
wavelet centered at $t_{0}$. Similarly the bandwidth  of the
wavelet is
\begin{equation}\label{delomega}
\Delta
\omega\equiv\sqrt{\frac{\int_{-\infty}^{\infty}(\omega-\omega_{0})^{2}|\varphi(\omega)|^{2}
d\omega}{\int_{-\infty}^{\infty}|\varphi(\omega)|^{2} d\omega}}.
\end{equation}
This mother wavelet is then used to build a set of daughter
wavelets by translating $\varphi(t)$ in time, and by dilating or
contracting $\varphi(t)$, which not only adjusts the mean
frequency but also the spread of the daughter wavelet.

Consider the case, when the mother wavelet is scaled by $s$. The
Fourier transform of $\varphi(t/s)$ is $|s|\varphi(s\omega)$. The
RMS duration becomes $\Delta t(s)=|s|\Delta t$ and the
corresponding RMS bandwidth is $\Delta\omega(s)=\Delta\omega/s$.
It implies $\Delta t(s)\Delta\omega(s)=\Delta t\Delta\omega$,
which is independent of the scaling
 parameter $s$. Also, the RMS bandwidth is not affected by
 the translation parameter $\tau$ because translating a function does
 not affect the magnitude of its FT.
 For our chosen mother wavelet $\varphi(t)$, Eqs.~(\ref{delt}) and (\ref{delomega}) yield $\Delta t=\Delta\omega=1/\sqrt{2}$
 so that $\Delta
\omega\Delta t=1/2$, which is independent of scaling and
translation. Thus the uncertainty of time localisation is
accompanied by an increase in the uncertainty of frequency
localisation or vice versa.

Referring to Fig.~\ref{ryd3}, we note that several patches appear
in a rectangular array on the time-frequency plane. Each patch is
centered about a particular frequency and a particular time. To
help us understand the occurrence of these patches, we undertake
an analytical approach as described below. Recall that the
absolute square of the autocorrelation function is given by
Eq.~(\ref{auto20}). Writing $y=(t-\tau)/s$, the CWT of $f(t)$ can
be written as
\begin{equation}
T(\tau,s)=\sqrt{s}\int_{-\infty}^{\infty} f(ys+\tau)\phi^{*}(y)dy.
\label{tf1}
\end{equation}
where $\phi(y)$ is the Morlet with shifted time $\tau$ and scaled
by $s$:
\begin{equation}
\phi(y)=\pi^{-1/4} e^{i\omega_{0} y} e^{-y^{2}/2}.
\end{equation}
Substituting in Eq.~(\ref{tf1}) and performing the integration
over $y$, we get
\begin{equation}
T(\tau,s)=\sqrt{2\pi s}\sum_{n,m}|c_{n}|^{2} |c_{m}|^{2}
\pi^{-1/4}
e^{-iE_{nm}\tau/\hbar}e^{-(\omega_{0}+sE_{nm}/\hbar)^{2}/2}.
\label{tf3}
\end{equation}
Maximum values of $T(\tau,s)$, corresponding to a particular scale
parameter $s$ should occur whenever the factor
$e^{-(\omega_{0}+sE_{nm}/\hbar)^{2}/2}$ approaches unity. This
gives rise to a constraint:
\begin{equation}\label{tf4}
\omega_{0}=-sE_{nm}/\hbar.
\end{equation}
Since the distribution functions $\vert c_n\vert^2$ and $\vert
c_m\vert^2$ are peaked about $\bar{n}$, the central frequency of
each frequency band in Fig.~\ref{auto1}(b) can be obtained by
setting $n=\bar{n}$ and $m=\bar{n} +p$, where $p$ is an integer.
Since $\omega_0$ is positive, we must insist that $p$ is a {\it
positive} integer. Substituting in Eq.~(\ref{tf4}) and using the
quadratic approximation ~(\ref{phase1}) for the energy
eigenvalues, we immediately get
\begin{equation}\label{new1}
\omega_{0}=\frac{2\pi s
p}{T_{\mathrm{cl}}}\left(1-p\frac{T_{\mathrm{cl}}}{T_{\mathrm{rev}}}\right)
\end{equation}
Since the scale parameter $s$ is related to frequency $f$ by the
relation $s=\omega_0/(2\pi f)$ and $T_{\mathrm{cl}}\ll
T_{\mathrm{rev}}$, we finally obtain the simple formulae
\begin{equation}\label{freq}
f_p=\frac{p}{T_{\mathrm{cl}}},\;\;\;s=\frac{\omega_0}{2\pi f_p}.
\end{equation}
The above formula correctly predicts the central frequencies
around which the spectral components are clustered in the
frequency plane as was shown earlier in Fig.~\ref{auto1}(b). Note
that each horizontal band in Fig.~\ref{ryd3} has a spread about
its central frequency. We will now show that the terms
corresponding to these frequencies add up coherently in the
expression for $T(\tau,s)$ at a particular time $\tau$ given by
\begin{equation}\label{frac}
\tau=\frac{k}{2p} T_{\mathrm{rev}}
\end{equation}
where $k$ is an integer.

Coherent addition of terms would require that the phase factor
$\exp(-i E_{nm}\tau/\hbar)$ be the same for these terms. That is,
for arbitrary values of $n$ and $n'$, one should be able to
satisfy the condition
\begin{equation}\label{new2}
\tau E_{n,n+p}=\tau E_{n',n'+p}+2\pi N
\end{equation}
where $N$ is an integer number. Using the approximation
~(\ref{phase1}), it is now easy to show that $\tau$ is indeed
given by the expression ~(\ref{frac}). An equivalent and simpler
way of deriving this result is to insist that the phase factor
$\exp(-i\tau E_{n,n+p})$ is independent of $n$.

The expression ~(\ref{frac}) gives us the time instants at which
$\vert A(t)\vert^2$ is peaked. Specifically, it tells us which
frequency beats occur at what times. The lowest frequency beats
$f_1$ occur for $p=1$ at times $t/T_{\mathrm{rev}}=1/2$, $1$,
$3/2$, $2$,.... Similarly, the frequency beats $f_2$ corresponding
to $p=2$, occur at $t/T_{\mathrm{rev}}=1/4$, $1/2$, $3/4$,
$1$,.... Thus the patches appearing in the bottom row give the
transition frequency between any two consecutive levels ($p=1$),
the ones on the next row are the transition frequencies
corresponding to $p=2$ and so on.  In this way, frequency bands,
depicted in Fig.~\ref{ryd3}, trace out the signature of fractional
revivals.

Thus interestingly, in time-frequency plane, one can find both
time and frequency information from the two time scales
$T_{\mathrm{cl}}$ and $T_{\mathrm{rev}}$ by using the expressions
~(\ref{frac}) and ~(\ref{freq}). In fact, these two expressions
provide the location of each patch in the time-frequency plane. As
an example, let us consider the fourth maximum or patch in the
third harmonic appearing in Fig.~\ref{ryd3}. Here, the third
harmonic corresponds to $p=3$, so the frequency information
corresponding to this patch can be obtained from Eq.~(\ref{freq}),
which is $f_3=1.457\times 10^{-8}$ a.u., as shown in
Fig.~\ref{ryd3}. The corresponding value for time can be obtained
from Eq.~(\ref{frac}). In this case, $k=4$, and $p=3$, so the
patch will appear at $t=\frac{2}{3}T_{\mathrm{rev}}$.

\begin{figure}[htbp]
\centering
\includegraphics[width=4.7in,height=5.8in]{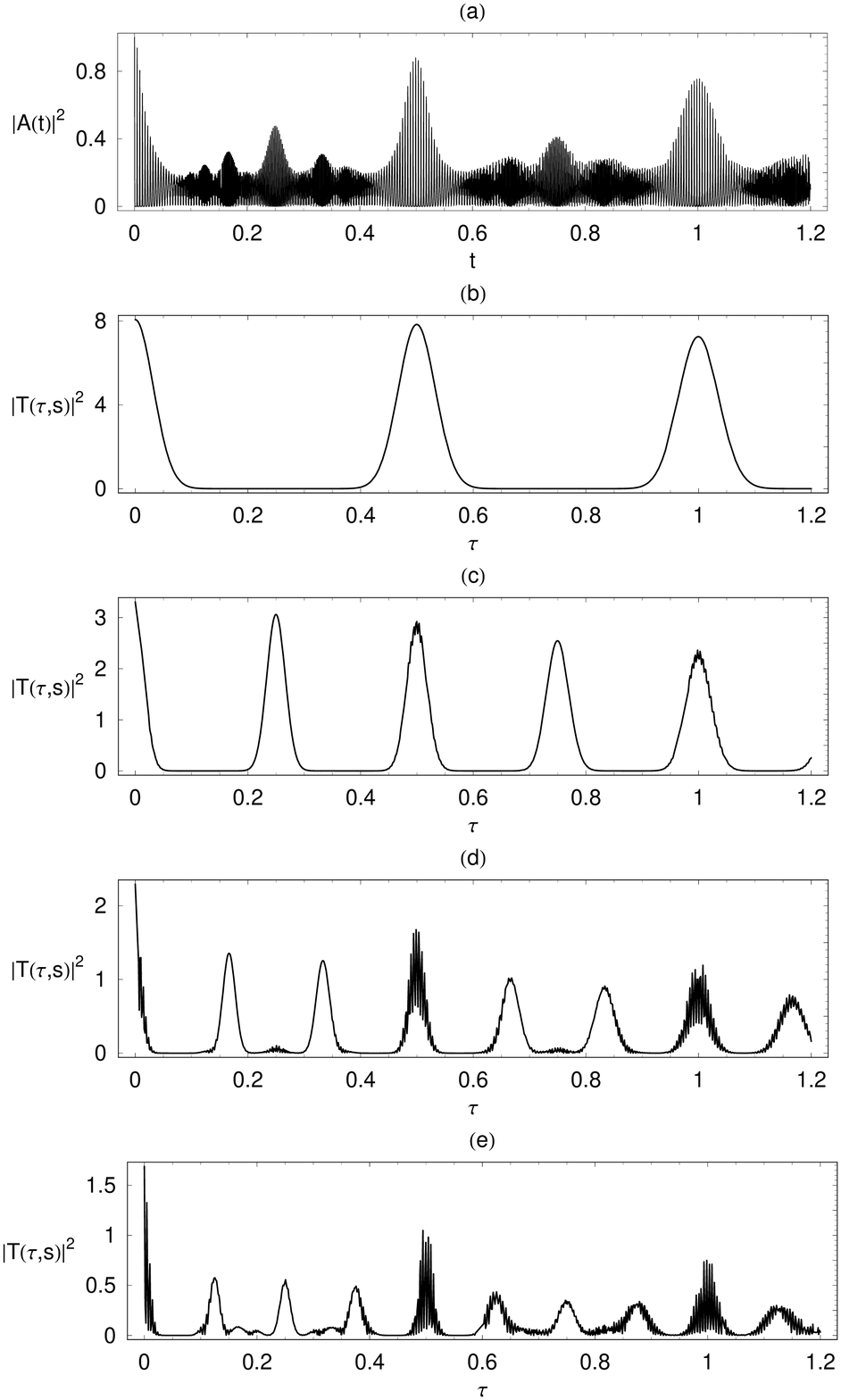}
\caption{(a)Autocorrelation function of Rydberg atomic system with
$\sigma=2.5$, (b), (c), (d)  and (e) show $|T(\tau,s)|^2$ for
$s=1.96,\;0.97$,\;$0.65$ and $0.48$ (in unit of $10^8$)
respectively. Times are scaled by revival time and $|T(\tau,s)|^2$
is scaled by $10^{6}$.} \label{tfr2d}
\end{figure}

\begin{figure}[htbp]
  \includegraphics[width=2.6in,height=1.95in]{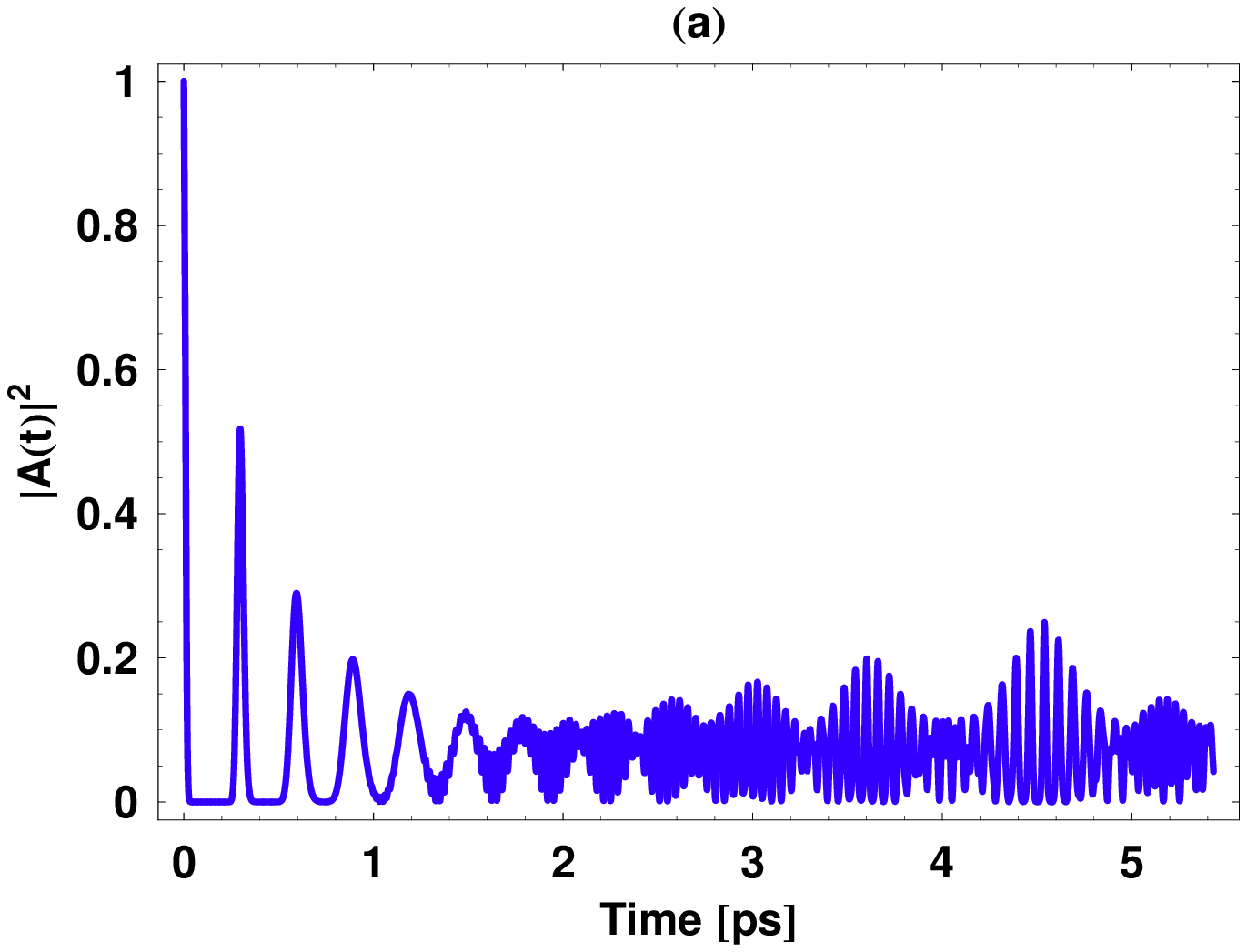}\;\;\;
  \includegraphics[width=2.4in,height=2.in]{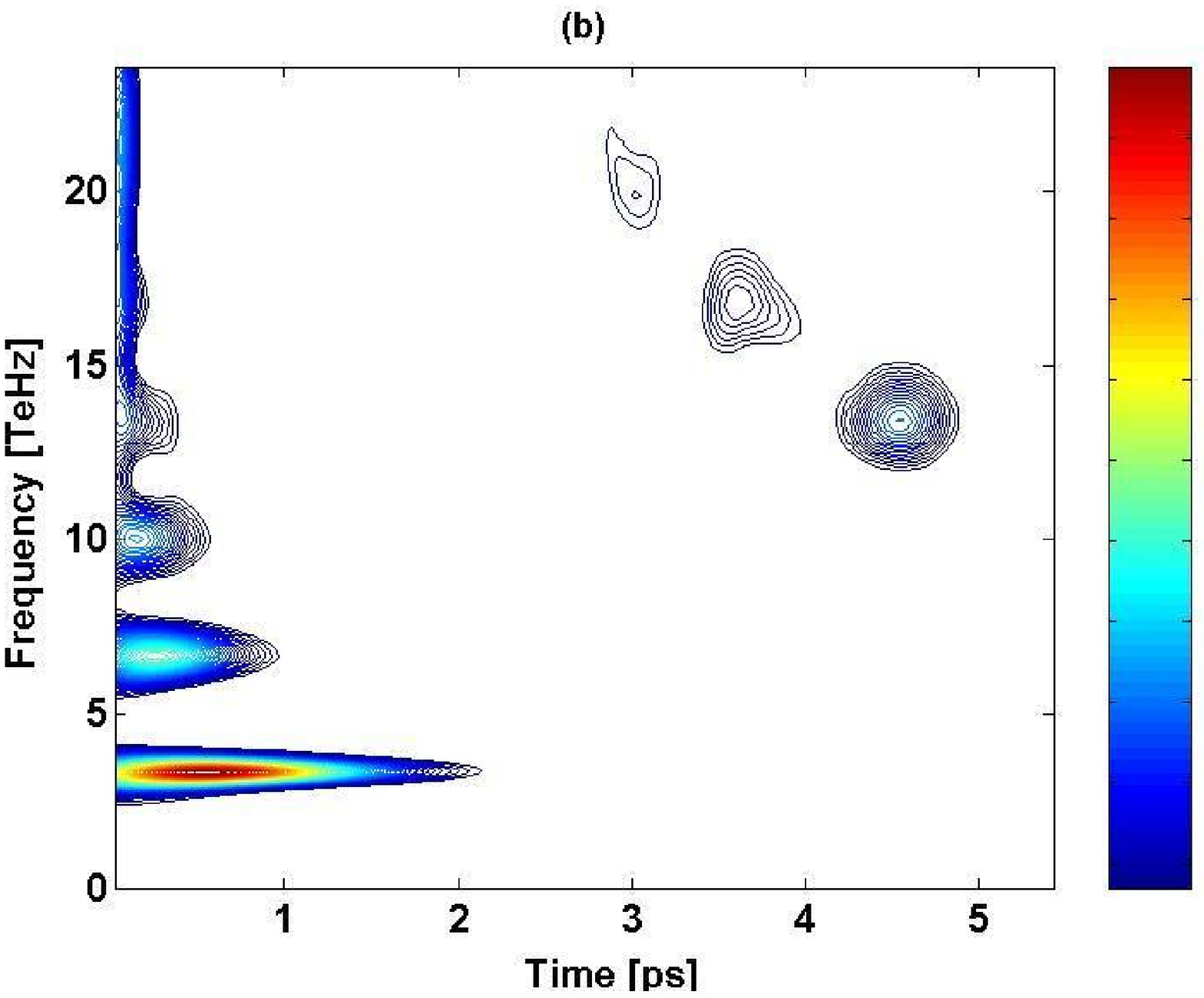}
\caption{(Color online) (a) A short-lived time series of
autocorrelation function and (b) its wavelet based time-frequency
representation.} \label{short}
\end{figure}

\section{Advantages of the Time-Frequency Representation}

It is seen that the wavelet based time-frequency representation
provides localisation in both time and frequency. Higher order
fractional revivals and their localisation in time are clearly
manifest in the time-frequency plane. We showed how
$T_{\mathrm{rev}}$ and $T_{\mathrm{cl}}$ are themselves sufficient
to explain the time-frequency plane completely.

 Note also that the square of the autocorrelation function, plotted as a time
series, does not resolve fractional revivals unambiguously. More
precisely, the order of the fractional revival cannot always be
determined. In contrast, the time frequency representation
introduces a parameter $p$ through Eqs.~(\ref{freq}) and
(\ref{frac}) that clearly separates out the fractional revivals in
a rectangular array on the time-frequency plane. In
Fig.~\ref{tfr2d}, we show how the corresponding $s$ values filter
out fractional revivals from the complicated plot of the
autocorrelation function. For $p=1$, one can find $s$ using
Eq.~(\ref{freq}). This specific $s$ value can filter out the
signature of fractional revivals as shown in Fig.~\ref{tfr2d}(b).
This is also true for the higher order harmonics as shown in
Fig.~\ref{tfr2d}(c), (d) and (e) for $p=2$, $p=3$ and $p=4$
respectively. These specific values of $s$, filter out the
signature of the corresponding higher order fractional revivals.

The time-frequency representation can be useful for a wave packet
that decays before its revival time. Although the short-time
evolution of $\vert A(t)\vert^2$ can still be used to estimate
$T_{\mathrm{cl}}$, no information can be gained about
$T_{\mathrm{rev}}$ if the energy spectrum of the system is {\it
not} known. However, as long as the wave packet survives long
enough for some patches to occur in the scalogram ( see
Fig.~\ref{short}), one can use Eq.~(\ref{frac}) to estimate
$T_{\mathrm{rev}}$. What makes this possible is the better
resolution available in the time-frequency plane for the detection
of fractional revivals.

\section{Conclusion}

In conclusion, we have made use of the continuous wavelet
transform to demonstrate the time-frequency representation of
autocorrelation function for the wave packet dynamics of a Rydberg
wave packet. An analytical approach is provided to interpret the
time-frequency plane and explain our numerical observations. We
have shown that the time-frequency representation not only
provides a complementary method of analyzing fractional revivals,
it is a better tool in resolving fractional revivals. Finally, it
is shown that the time-frequency representation may be able to
extract information about the revival dynamics of a short-lived
system even if the system decays before reaching its revival time.
\ack
We thank Dr. P. K. Panigrahi for fruitful discussions and
valuable comments.

\end{document}